\documentclass[a4paper,12pt]{article}
 \usepackage[latin1]{inputenc}
\usepackage[T1]{fontenc}
\usepackage{amsmath}
\usepackage{amsfonts}
\usepackage{amssymb}
\usepackage{color}
\usepackage{graphicx}

 \newcommand{\be}{\begin{equation}}
	 \newcommand{\ee}{\end{equation}}
	 \newcommand{\ba}{\begin{eqnarray}}
		 \newcommand{\ea}{\end{eqnarray}}
		 
		   \newcommand{\bea}{\begin{eqnarray}}
			 \newcommand{\eea}{\end{eqnarray}}

\begin{document} \title{Lifshitz holography with a probe Yang-Mills field}
\author{ Fidel A. Schaposnik\thanks{Also at CICBA}\; and
Gianni Tallarita \\ \vspace{0.2 cm} \\
{\normalsize \it Departamento de F\'\i sica, Universidad Nacional de La Plata}\\ {\normalsize \it Instituto de F\'\i sica La Plata}\\ {\normalsize\it C.C. 67, 1900 La Plata, Argentina}}

\date{\hfill}

\maketitle
\begin{abstract}
Taking
  as a probe an $SU(2)$ gauge field with Yang-Mills action in a $3+1$ dimensional Lifshitz black hole background, we use
   the gauge/gravity correspondence to discuss  finite temperature effects in the dual theory defined on the boundary. In order to test the dependence of results on the anisotropic scaling exponent $z$  we  consider two analytical black hole solutions with    $z =2$ and $z= 4$.  Apart from solving the equations of motion in the bulk  using a numerical approach, we also apply an analytical approximation  allowing the  determination of the phase transition character, the critical exponent and  the critical   temperature behavior as a function of $z$.
\end{abstract}

Models with anisotropic scaling were   introduced  in condensed matter physics more than thirty years ago in order to discuss tricritical points (see \cite{Diehl} and references therein). They are at present actively investigated in the context of gravitational theories in which space-time anisotropic scaling improves the short distance behavior (see \cite{Hori} and references therein).
A link between these two issues was established by Kachru, Liu and Mulligan \cite{Kachru}
within the framework of the gauge/gravity correspondence by searching gravity duals of non-relativistic   quantum field theories. Studying
the equations of motion of Einstein gravity with negative cosmological constant coupled to $p=1$ and $2$-forms  a solution  was found in \cite{Kachru} with the metric taking the form
\be
ds^2 =  {L^2} \left (-  r^{2z} dt^2 + r^2 d\vec x^2 + \frac{dr^2}{r^2} \right)
\label{cero}
\ee
where $0 < r <\infty$, $d\vec x^2 = dx_1^2 + \ldots dx_n^2$,  $L$ is the radius of curvature of the geometry and $z\geq 1$.  Metric (\ref{cero})
is invariant under anisotropic scaling of space-time coordinates
\be
t \to  \lambda^{z} t \; , \;\;\;\; \vec x \to \lambda \vec x \; , \;\;\;\;  r \to \frac{r}\lambda
\label{uno}
\ee
with $z$ playing the role of the dynamical critical exponent
\cite{Hori}. The coordinates' inverse length dimensions are: $[t] = -z, [r] = +1, [x]=[y]= -1
$.
%
Taking eq.\,(\ref{cero}) as a  background metric,  the authors in ref. \cite{Kachru} extended the gauge/gravity duality to the case of models with anisotropic scaling and explored the boundary observables dual to free  scalar fields in a $3 + 1$ dimensional bulk.

The finite temperature extension of the gauge-gravity duality requires to consider a black-hole bulk metric with line element
\be\label{gensol}
ds^2 = L^2\left(-{g_z(r)}{r^{2z}}dt^2+\frac{1}{g_z(r)r^2}dr^2+{r^2}(dx^2 + dy^2)\right)
\ee
where  $g_z$ vanishes at the horizon $r_H$.   Different black-hole solutions  with anisotropic scaling are available \cite{Taylor}-\cite{Bala} and a number of holographic studies have considered them  as a background with bulk Lagrangians including different fields: charged matter,    Abelian and non Abelian  gauge fields, and  massive Proca fields \cite{Bry}-\cite{Bu}.

Using the gauge/gravity correspondence we study in the present work finite temperature effects in the dual theory defined on the boundary. We take as a probe   an $SU(2)$ gauge  field $A_\mu$ with Yang-Mills action, this implying that the order parameter is a vector and that one should expect a strongly anisotropic result for conductivities (among the works cited above, solely ref.
\cite{Bu} has considered a vector order parameter).  In order to test the dependence of results on $z$ we shall  consider two analytical $3+1$ dimensional black hole solutions with different  $z$ values:  the $z=2$ black hole found in \cite{Bala} and the one presented in \cite{Bry} and \cite{Dehghani} for the $z=4$ case.

The $z=2$ black hole constructed in \cite{Bala} arises as
a solution of the equations of motion for a $3+1$ dimensional  gravitational theory with negative cosmological constant coupled to a massive vector field ${\cal A}_\mu$
 and a scalar field $\Phi$ without kinetic term. The action reads
 \be
 S_2 = \frac12\int d^4x (R - 2\Lambda) -\int d^4x \left(\frac14\exp(-2\Phi) {\cal F}_{\mu\nu}
 {\cal F}^{\mu\nu}
+ \frac{m^2}2 {\cal A}_\mu {\cal A}^\mu  +
\left(  \exp(-2\Phi) -1 \right)
\right)
\label{S2}
 \ee
The solution of the equations of motion corresponds to a metric with line element given by
eq.\,(\ref{gensol}) with
\be
g_{\small 2}(r) = 1-\frac{r_H^2}{r^2} \, .
\label{greevy}
\ee
Starting from an action in which  a Maxwell field $A_\mu$  is coupled to gravity but not directly to the massive vector field,
a charged
  $z=4$ flat horizon black-hole solution  was presented in refs. \cite{Bry},\cite{Dehghani}. The action
  takes in this case the form
  \be
S_4 = \frac12\int d^4x (R - 2\Lambda) -\int d^4x \left(\frac14{ F}_{\mu\nu}
 {F}^{\mu\nu}+ \frac14{\cal F}_{\mu\nu}
 {\cal F}^{\mu\nu}
+ \frac{m^2}2 {\cal A}_\mu {\cal A}^\mu \right)
 \label{S4}
\ee
with the black holes function $g_{4}$ given by
\be
\label{chargedsol}
g_{4}(r)=1-\frac{ Q^2}{8r^4} \, .
\ee
where $Q$  an integration constant related to the Maxwell field.

The  black hole temperature  associated with (\ref{gensol}) is given by
\be
T_z=\frac{1}{\beta}=\frac{|g_{z}'(r_H)|r_H^{z+1}}{4\pi}
\label{become}
\ee
so that  for the   $z=2,4$   black holes described above one has
\be
T_{2} = \frac{r_H^2}{2\pi}  , \;\;\;\;\;\;\;\;\;\;  \;\;\;\;\;    T_{4} = \frac{Q^2}{8\pi} \;.
\label{temp}
\ee
Note that $[T_z] = z$.

As stated above, we take as a probe an $SU(2)$ gauge  field $A_\mu^a$ ($a=1,2,3$) in the
black hole background  (\ref{gensol}) with $g_z(r)$ given by (\ref{greevy}) and (\ref{chargedsol}).
We take from here on $L=1$.
We start from the Yang-Mills action
\be S = -\frac{1}{4}\int d^4x\,\sqrt{|g|}\;  \; F_{\mu\nu}^a F^{a\,\mu\nu}
 \label{1} \ee
The field strength   $F^a_{\mu \nu}$   ($a=1,2,3$)  is defined as
\be
F^a_{\mu \nu} = \partial_\mu A_\nu^a - \partial_\nu A_\mu^a +
\varepsilon^{abc}A_\mu^b A_\nu^c \label{6}
\ee
We have taken the gauge coupling constant equal to one.
The equations of motion read
\be
\frac{1}{\sqrt{-g}}\partial_\mu(\sqrt{-g}F^{a \mu\nu})-\epsilon^{abc}F^{b\nu\mu} A^c_{\mu} =0.
\label{eqs}
\ee
In order to solve these equations we shall consider the  ansatz proposed in \cite{Gubser} for a relativistic non-Abelian gauge theory defined in an asymptotically AdS space-time
\be
A =\phi(r)\tau^3 dt+\omega(r)\tau^1dx
\label{13}
\ee
where $\tau^a$ are Pauli matrices. The gauge field inverse length dimensions are $[\phi] = z$ and $[\omega] = 1$.
It will be convenient to introduce the dimensionless variable  ${u} = r_H/r$, so that
the horizon is located at $u = 1$ and the asymptotic  boundary at $u=0$. In terms of
this variable, and inserting the black hole metric (\ref{gensol}),  eqs.(\ref{eqs}) reduce to
\be\label{eom1}
\phi''+\frac{z-1}{ {u}}\phi'-\frac{1}{r_H^2g( {u})}\phi\omega^2=0
\ee
\be\label{eom2}
\omega''+\frac{ {u}^{z-1}}{g( {u})}\partial_{ {u}}({u}^{1-z}g( {u}))\omega'+
\frac{{u}^{2z-2}}{r_H^{2z}g({u})^2}\omega\phi^2=0.
\ee

Let us discuss appropriate  conditions for the gauge field components.
The consistent conditions at the $u = 1$ horizon are
\ba\label{horizon}
&&\phi\sim\phi_1(1-u)+...
\\
&& \hspace{5.3 cm}    u \to 1\nonumber
\\
&&\omega\sim \omega_H + (1-u)^2\omega_1+...
\label{horizon2}
\ea
 Concerning the boundary $u = 0$, one has
\ba
&& \phi   \sim \mu+ \rho \ln(u)+... \\
&& \hspace{4 cm} z = 2 \,, \;\;  u \to 0 \nonumber
\\
&& \omega  \sim \omega_0+ \Omega u^{2}+...
\label{notienenombre}
\ea
\ba
\label{bdry}
&&\phi  \sim \mu + \rho u^{2-z}+...
\\
&& \hspace{4 cm}  z > 2 \,, \;\; u \to 0 \nonumber
\\
&&\omega   \sim \omega_0+\Omega u^{z}+...
\label{19}
\ea

According to the gauge/gravity correspondence $\mu$ will be identified with  the chemical potential and $\rho$ with  the total charge density  in the
dual theory defined on the boundary.

The general   solution for $\phi$ with $z=2$ in the normal phase    takes the form
\ba
\phi_n &=& \mu_n +\rho\ln\left(u\right) \nonumber\\
\omega &=&0
\ea
Using the horizon condition $\phi(1)=0$, we have that
\be
\mu_n = 0
\ee
so that the chemical potential of the normal phase vanishes.
In contrast, for the $z=4$ normal phase one has
\ba
\phi_n &=& \rho\left(1 - \frac{1}{u^2}\right) \nonumber\\
\omega &=&0
\label{25}
\ea
and hence the chemical potential of the $z=4$ normal phase is non-vanishing, $\mu_n = \rho$.

In the   $z=1$ relativistic case the divergencies of the action at the boundary are eliminated by adding counterterms.  New divergent terms arise for $z \geq 2$ but  taking a fixed charge
density $\rho$ as boundary condition make these terms temperature independent   \cite{HPST}. We thus adopt   this  natural choice in what follows.  If, as it happens in the $z=1$ case \cite{Gubser},  ansatz (\ref{13})  for a $z>1$
 theory can be related to a  an holographic p-wave superconductor, the order parameter should then  be $\Omega$. The necessary requirement for   $\Omega$ to be unsourced forces the choice of vanishing $\omega_0$
  in eq.\,(\ref{19}) or eq.\,(\ref{25}). The divergencies of the action in the normal $\omega = 0$ phase and the superconducting $\omega \neq 0$ one  coincide leading to a finite free energy difference, as we shall see below.

We shall now proceed to calculate
the free energy  ${\cal F}$, related to the Euclidean on-shell action according to
 \be
 {\cal F} = \left. T_zS_E\right|_{on\,shell}
 \label{esta}
 \ee
 Before proceeding to the   Wick rotation of the action we insert the ansatz
 (\ref{13}) in eq.\,\eqref1
 \be\label{s1}
  S  = -\frac{V}{2T_z}\int du \frac{1}{u^{3+z}}
  \left(-r_H^{2-z}u^{2z+2}(\phi')^2-
  r_H^{-z}\frac{u^{2z+2}}{g_z(u)}\omega^2\phi^2+r_H^z u^4(\omega')^2g_z(u)\right)
\ee
 where $V$ is the two dimensional boundary spatial volume.

 We start with the $z=2$ case.
Integrating by parts eq.\,\eqref{s1}  and using the equations of motion  we get  
\be
\frac{T_2}{V}S  = \frac{1}{2}\left[(u\phi\phi')|_{u=\epsilon}- \frac{r_H^2g(u)}{u}\omega'\omega|_{u=\epsilon}\right]-\frac{1}{2}\int du\frac{u}{r_H^2g(u)}\phi^2\omega^2.
\ee
Here $\epsilon$ is a cut-off which will be put to zero at the end of the calculations.
As discussed above, we choose to work in the canonical ensemble and hence we add a boundary term to the action \cite{Skenderis}
\be
-\frac{1}{2}\int dt d^2x   \sqrt{-g}A_\mu F^{u\mu}|_{u=\epsilon}=\frac{V}{2T_2}\left[(u\phi\phi')|_{u=\epsilon}-\frac{r_H^2g(u)}{u}
\omega'\omega|_{u=\epsilon}\right].
\ee
After a Wick rotation, using eq.\,\eqref{esta} and  the boundary behavior  of the gauge field the free energy density at fixed charge takes the form
\be
\frac{\mathcal{F}}{V} = -\rho\mu+\frac{1}{2}\int du\frac{u}{r_H^2g(u)}\phi^2\omega^2 + \frac{1}{2}\rho^2\ln\left(u\right)|_{u=\epsilon}
\ee
  The logarithmic divergent term in the r.h.s. will play no role
when comparing the free energies of the solutions with $\omega \ne 0$ with  that of the normal $\omega = 0$ case  which has    the same divergent term so that
one ends with
\be
\frac{\Delta\mathcal{F}}{V} = \frac{\mathcal{F} - \mathcal{F}_n}{V}   = -\rho\mu+\frac{1}{2}\int du\frac{u}{r_H^2g(u)}\phi^2\omega^2  \; . \;\;\;\;\;\;\;\;\;\;\;\;  z=2
\label{31}
\ee
Proceeding in the same way in the $z = 4$ case, we find
\be
\frac{\Delta\mathcal{F}}{V} = -\rho(\mu -\mu_n)+\frac{1}{2}\int du\frac{u^3}{r_H^2g(u)}\phi^2\omega^2 \; , \,\,\;\;\;\;\;\;\;\;~~\;\;\;\;\;\;\;\;\;  z=4
\label{32}
\ee
where $\mu_n$ is the chemical potential of the normal phase.

Before discussing the numerical solutions of equations (\ref{eom1})-(\ref{eom2})  we shall develop an analytic approach which   allows to calculate the critical temperature and the behavior of the order parameter with remarkable accuracy. The method is based in a proposal presented in ref. \cite{Gregory} which consist  in obtaining solutions in close form by imposing conditions of continuity and smoothness  at a point $u_m$ intermediate between the boundary ($u=0$) and the horizon ($u=1$). Originally $u_m$ was arbitrarily chosen to be 1/2 and rather good results in comparison with more involved numerical methods were obtained.  As discussed in \cite{Bellon} the agreement stems from rather elementary considerations on perturbation of Schr\"odinger-like equations. We here extend the method in order to determine   $u_m$ from a simple free energy argument and in this way,  the method turns out to be a powerful tool to study the behavior of the system as a function of $z$.

In
practice, we shall consider  expansions of the fields   near $u=1$ and $u=0$ and determine their leading orders
coefficients by connecting the expansions at $u=u_m$.
We start from the case $z=2$. For the
solution near the horizon ($u=1$) we have, up to second order
  in the expansions of the fields we call $ \omega^h(u)$ and $\phi^h(u)$,
\begin{align}
\omega^h(u)& =  \omega_0^h +  \omega_1^h\, (u-1) + \frac{1}{2}\,  \omega_2^h\,
(u-1)^2
\nonumber\\
 \phi^h(u)& =   \phi_0^h + \phi_1^h\, (u-1) + \frac{1}{2}\,    \phi_2^h\,
(u-1)^2 \label{expansion0}
\end{align}
where $ \omega_i^h,  \phi_i^h, \,  $
are constants to be determined. The superscript $h$ indicates that the expansion is performed near the
horizon.
Now, conditions (\ref{horizon}) at $u = 1$ imply that
\ba
&&  \phi_0^h = 0 \; , \;\;\;\; \phi_2^h= \frac14 \phi_1 \left(2 + \frac{  \omega_H^2}{r_H^2}\right)
\\
&&  \omega_1^h =0 \; , \;\;\;\;  \omega_2^h = -\frac1{16} \frac{ \phi_1^2\omega_H}{r_H^4}
\ea
We now insert these relations in eq.~(\ref{expansion0}) and match the expansions of $\omega$ and
${\phi}$  and their derivatives  at $u = u_m$. From this we get
\ba
\phi_1 &=&  -\frac{4r_H^2}{\sqrt{1-u_m}} \; , \;\;\;\; \;\;\;\; \;\;\;\; \Omega  = \frac{\omega_H}{u_m}
\\
\omega_H &=& \left( 2r_H^2\frac{2 - u_m }{1 - u_m} - \frac{\rho}{2u_m\left(1 - u_m\right)^{1/2}}\right)^{1/2}
\ea
At this point we can write $r_H$ in terms of the temperature $T$ using eqs.(\ref{temp})
\be
\Omega = \frac1{u_m} \left(4\pi T_2\frac{2 - u_m }{1 - u_m} - \frac{\rho}{2u_m\left(1 - u_m\right)^{1/2}}\right)^{1/2}
\ee
Determination of the  point at which the order parameter $\Omega$  vanishes leads to the critical temperature
\be
T^{c}_{2} = \frac{1}{8\pi }\, \frac{(1-u_m)^{1/2}}{u_m\left(2-u_m\right)} \rho.
\label{40}
\ee
One can also infer the temperature dependence of the condensate close the phase transition
\be
\Omega = {\cal N}_2(u_m)  \left({4\pi T_2^{c}}\right)^{1/2}\left(1-\frac{T}{T_2^{c}}\right)^{1/2}
\label{41}\ee
\be {\cal N}_2(u_m) =  \frac{1}{u_m}\left(\frac{{2-u_m}}{{1-u_m}}\right)^{1/2}
\ee
Similar calculations with $z=4$ yield, using $Q^2 = 8 r_H^4$
\ba
\phi_1 &=& -\frac{Q^2 }{(1 - u_m/{2})^{1/2} (1 - u_m)^{1/2}} \; , \;\; \;\;\;
 \;\;\; \Omega = \frac{\omega_H}{u_m^3 (2 - u_m)}
\\
\omega_H &=& - 2^{1/4}\left(  {Q} \frac{4 - 3 u_m}{u_m - 1}  +
 \frac{4\rho}{Q^2}\frac{(2 - u_m)^{1/2}}{u_m^3 (1 - u_m)^{1/2}}
\right)^{1/2}
\ea
which defines the critical temperature as
\be
T_4^c= \frac{(2-u_m)^{1/2}(1 - u_m)^{1/2}}{2^{5/2}\pi u_m^3(4 - 3 u_m)} \rho
\label{46}
\ee
%
Finally for the behaviour of the order parameter near the critical temperature we obtain
\be
{\Omega}= {\cal N}_4(u_m) \left({16\pi T_4^{c}}\right)^{1/4}
\left(1-\frac{T}{T_4^c}\right)^{1/2} \label{47}
\ee
\be
{\cal N}_4(u_m) = \frac1{u_m^3}\frac{(4 - 3u_m)^{1/2}}{(2 - u_m)(1 - u_m)^{1/2}}
\ee

We then see that both for $z=2$ and $z=4$ the behavior of $\Omega$  near the critical point reveals  a typical scenario of a second order phase transition, with an ordered
phase $\omega \ne 0$ for $T < T^c_z$  in agreement
with  the results in the most diverse relativistic models explored using the gauge/gravity duality, with critical exponents coinciding with those obtained within the mean field approximation,
independently of the choice of $u_m$.

To confirm the results obtained above  we have still to compare the free energy  associated to the solution we have found with that for the disordered (normal) phase   which corresponds to
$\omega = 0$. If the difference of free energies $\Delta{\cal F}$,
is negative below the critical temperature then a phase with non-vanishing order parameter will be preferred for $T < T_z^c$. This fact will allow us to determine $u_m$ as a function of $\rho$, from minimization of
$\Delta{\cal F}$  written in terms of expansion (\ref{expansion0}) from the horizon to $u_m$ and of expansions \eqref{bdry}-\eqref{notienenombre} from the boundary to $u_m$,
\be
\frac{\Delta{\cal F}}{V} = -\rho(\mu - \mu_n) +
\frac{1}{2}\int_0^{u_m}  du\frac{u}{r_H^2g(u)}(\phi^b\omega^b)^2  +\frac{1}{2}\int_{u_m}^1 du\frac{u}{r_H^2g(u)}(\phi^{h}\omega^{h})^2
\label{in}
\ee
where $\phi^b$ and $\omega^b$ are given by \eqref{notienenombre} for $z=2$ and \eqref{19} for $z=4$. Note that we have not included the divergent term
in \eqref{in} since we are working at fixed $\rho$ and hence such term is $u_m$ independent. Minimization
of eq.\,\eqref{in} gives a solution for $u_m$  which, inserted in eqs. \eqref{40} and \eqref{46} gives the following critical temperature coefficients
\be
T_{2}^c = 0.022 \,\rho   , \;\;\;\;\;\;\;\;\;\;  \;\;\;\;\;    T_{4}^c = 0.025 \, \rho
\label{tempcritsanal}
\ee
We will confirm below this  scenario and compare these results with those obtained  by solving the
 equations of motion numerically.  Before doing this let us note that the critical temperature obtained analytically increases when changing from the $z=2$ to the $z=4$ system. To determine whether this is a general behavior for arbitrary values of $z$ is relevant in connection with the theory of Fermi liquids \cite{HPST}. To analyze this issue in more general terms one can take  for illustrative purposes the following black hole function
\be
g(u;z) = 1 - u^z
\label{estaforma}
\ee
which includes the actual Lifshitz $z=2$ and $z=4$ black hole solutions studied previously. From eq.\,(\ref{become}) one can  write $r_H$ in terms of $z$ and $T$ and then, using the   analytical approach one can confirm     that,
for black holes of the form (\ref{estaforma}), $T^c_z$ is a growing function of $z$ for $z \geq 2$ for any choice of $u_m$.

We now proceed to solve the equations of motion numerically. The strategy  is the following: the solutions are searched as functions of the parameters $\omega_H$ and $\phi_1$ at the horizon with vanishing constant term for $\phi$    and with general non-vanishing $\omega_0$ at the boundary (see eqs. \eqref{horizon}-\eqref{notienenombre}). Then the numerical system is solved searching  possible values of $\phi_1$ at the horizon for which $\omega_0$ vanishes.
 In this way we have  obtained a set of  solutions for different field values at the horizon. The existence of several solutions  satisfying the appropriate boundary conditions, each one
 corresponding to a different value of $\phi_1$, is a phenomenon  already present in the relativistic case \cite{Gubser}. For increasing values of $\phi_1$ the solution for for $\omega$ has an increasing number
 of nodes $n$. Now,   evaluation of the free energy shows that it increases with the number of nodes and hence we conclude that solutions with $n\geq 1$ are energetically disfavored  so that we shall
 solely discuss the zero-node solution.

Our numerical solution confirms the results found analytically: a finite temperature continuous symmetry
breaking phase  transition takes place both for $z=2$ and $z=4$.  As shown in Figure 1 the system condensates at a critical temperature $T_c$. The behavior near $T_c$ can be seen, by fitting the curve, to correspond to a second order transition with critical exponent $1/2$ as advanced by the analytical result, eqs.\eqref{41}-\eqref{47}. It should be stressed that profiles for $z=2$ and $z = 4$ are strikingly
resemblant. What distinguishes the two cases is the value of the critical temperatures:
\be
T_{2}^c = 0.023 \,\rho   , \;\;\;\;\;\;\;\;\;\;  \;\;\;\;\;    T_{4}^c = 0.031 \, \rho
\label{tempcrits}
\ee
Comparing these values with those obtained previously using the analytic approach eqs.\,\eqref{tempcritsanal} we find a remarkable agreement.

~

\centerline{\includegraphics[width=1.08\textwidth]{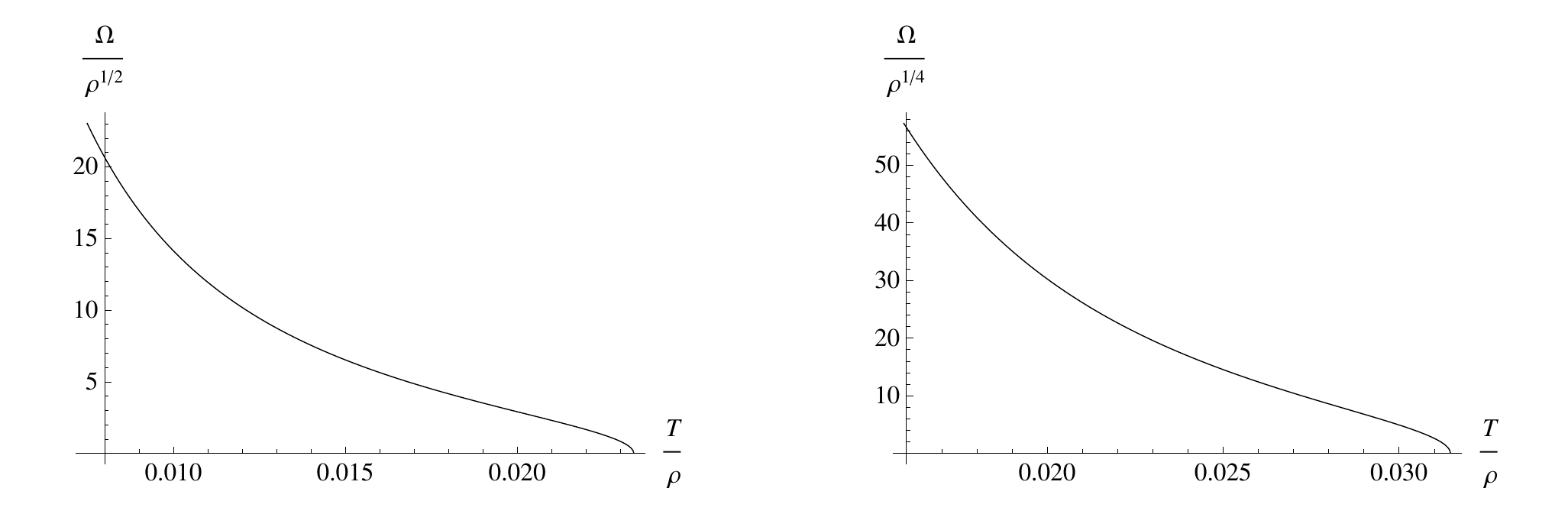}}

  \noindent Figure 1: The numerical result for the condensate as a function of temperature for the $z=2$ (left) and $z=$ (right) cases. The  condensate goes to zero as $(T -T^c)^{1/2}$ in both cases thus confirming the analytic results (eqs.\, \eqref{41},\eqref{47}).

  ~

  Note that at low temperature the condensates appear to diverge as a negative power of the temperature. This behavior was already encountered in the relativistic $z=1$ case, both for $s$-wave \cite{horowitz} and $p$-wave \cite{Gubser} holographic superconductors   and can be ascribed to the relevance of back-reaction when the
  condensate becomes too large so that the probe approximation is no more valid.   Using again eq.(\ref{estaforma}) as an illustration,  our analytical approach shows that
  the behavior of the condensate for $T$ small -in the range of validity of the probe approximation- is $\Omega \propto T^{-(z-2)/2z}$ for $z\geq 2$ independently of the choice of the matching point $u_m$.

 Using formul\ae ~\eqref{31}-\eqref{32} we have computed numerically the free energy difference between the ordered and disordered phases (see Figure 2)  confirming  that, both for $z=2$ and $z=4$, the ordered phase is preferred below the critical
 temperature $T^c_z$ whose values coincide with those given by \eqref{tempcrits}.

 ~

\centerline{\includegraphics[width= 1.04\textwidth]{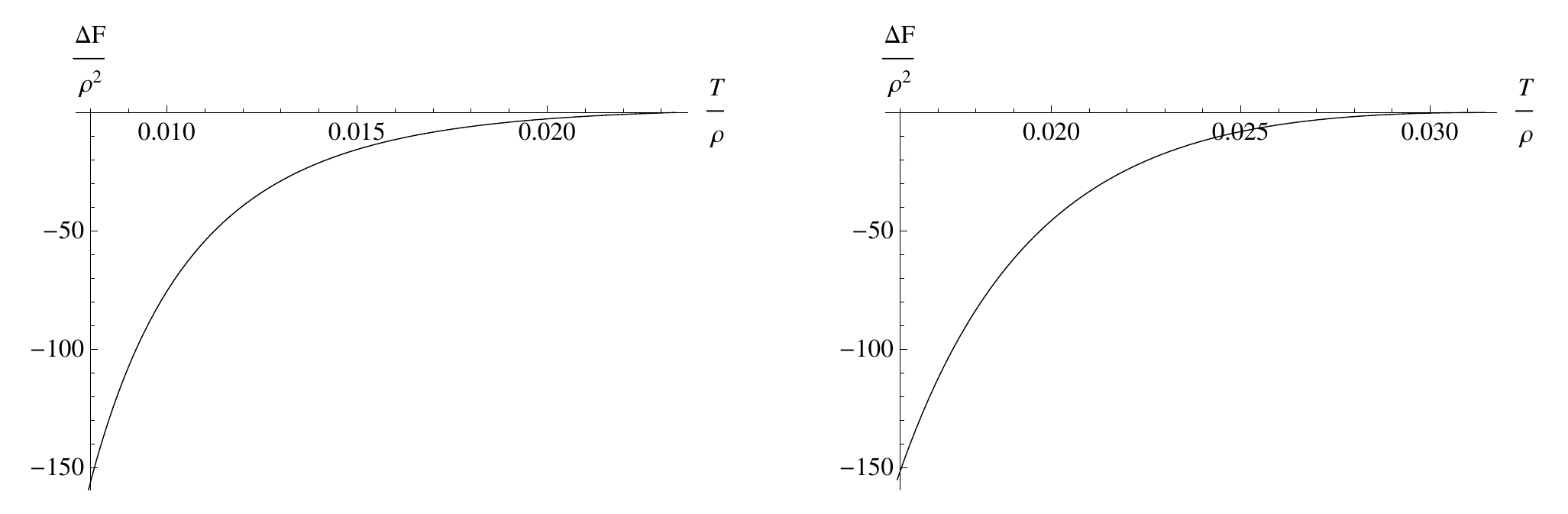}}

  \noindent Figure 2: The free energy difference between the condensed and the uncondensed phase
  as a function of temperature for the $z=2$ (left) and the $z=4$ (right) models

  ~

  Finally, we shall compute the electromagnetic response to small time dependent perturbations of  the electromagnetic fields
in the ordered  phase. To do this, we start from the gauge field ansatz  (\ref{13}) (that we shall denote $A_\mu^{ord}(u)$ for clarity) and following  \cite{Gubser} we consider the perturbation
\ba
A_\mu &=& A_\mu^{ord}(u) + a_\mu(u,t) \\
a_\mu dx^\mu &=& e^{-iw_ft}\left[(a_t^1\tau^1+a_t^2\tau^2)dt+a_x^3\tau^3dx+a_y^3\tau^3dy\right]
\label{ach}
\ea
with $w_f$   the   frequency associated to the perturbation.
The linearized Yang-Mills equations read
\be\label{linear}
\frac{1}{\sqrt{-g}}\partial_\mu(\sqrt{-g}\mathcal{F}^{\mu\nu a})-\epsilon^{abc}\mathcal{F}^{\nu\mu}_b A_{\mu c}-\epsilon^{abc}F^{\nu\mu}_b a_{\mu c}=0
\ee
where
\be
\mathcal{F}_{\mu\nu}^a=\partial_\mu a_\nu^a-\partial_\nu a_\mu^a-\epsilon^{abc}A_{\mu b}a_{\nu c}+\epsilon^{abc}A_{\nu b}a_{\mu c}.
\ee
Using eqs.\,\eqref{13},\eqref{ach} one finds four second order equations
\ba
\label{indep}
&&{a^3_y}''+\left(\frac{1-z}{u}+\frac{g'(u)}{g(u)}\right){a^3_y}'+\frac{w_f^2u^{2z-2}}{r_H^{2z}g^2(u)}a^3_y-
\frac{\omega^2}{r_H^2g(u)}a^3_y=0,
\label{ay}\\
&&
{a^3_x}''+\left(\frac{1-z}{u}+\frac{g'(u)}{g(u)}\right){a^3_x}'+\frac{u^{2z-2}}{r_H^{2z}g^2(u)}\left(-iw_f\omega a^2_t+w_f^2a^3_x-\omega\phi a^1_t\right)=0,
\label{x}\\
&&{a^1_t}''+\frac{z-1}{u}{a^1_t}'+\frac{\omega\phi}{r_H^2g(u)}a^3_x=0\label{at1}
\\
&&{a^2_t}''+\frac{z-1}{u}{a^2_t}'-\frac{\omega^2}{r_H^2g(u)}a^2_t-\frac{iw_f\omega}{r_H^2g(u)}a^3_x=0.
\label{at2}
\ea
and two first order equations
\be
iw_f{a^1_t}'+\phi {a^2_t}'-\phi{a^2_t}'=0
\ee
\be
iw_f{a^2_t}'-\phi {a^1_t}'+\phi' a^1_t-g(u)u^{2-2z}(\omega\partial_u-\omega')a^3_x=0.
\ee
Let us concentrate on the case $z=2$. The choice  of the electromagnetic perturbation should correspond to a wave traveling away from the conformal boundary at $u=0$ (an ``in-going'' wave). In the present case one has, near the horizon
\ba
a^3_y \!\!\!&=&\!\!\!  \alpha (1-u^2)^{-\frac{iw_f}{2r_H^2}}(1  + \ldots)
\; , \;\;\;  \;\;\; \;\;\;  \;\;\; \;\;\;  \;\;\;
a^3_x  = \beta (1-u)^{-i\frac{w_f}{2r_H^2}}(1+a_1(1-u)+ \ldots)
\\
a^1_t \!\!\!&=&\!\!\!\gamma(1-u)^{-i\frac{w_f}{2r_H^2}}(a_2(1-u)^2+ \ldots) \; , \;\;\;  \;\,
a^2_t =\delta(1-u)^{-i\frac{w_f}{2r_H^2}}(a_4(1-u) + \ldots)
\ea
with $\alpha,\ldots,\gamma$ dimensionful constants. At the boundary we have instead
\ba
a^3_y \!\!\!&=&\!\!\! a^3_{y(0)}+u^2a^3_{y(2)}+... \, ,  \;\;\; \;\;\;  \;\;\;  \;\;\; \;\,
a^3_x = a^3_{x(0)}+u^2a^3_{x(1)}+...
\\
a^1_t \!\!\!&=&\!\!\! a^1_{t(0)}+a^3_{t(1)}\ln\left(u\right)+...\, ,  \;\;\; \;\;\;  \;\;\;
a^2_t = a^2_{t(0)}+a^2_{t(1)}\ln\left(u\right)+...
\ea
where all coefficients $a_i$ can be determined as functions of $\omega$ and $\phi$ at the horizon once $w_f$ is specified.

  The conductivity can then be obtained using Ohm's law. Following  \cite{Gubser} for  the case of non-Abelian gauge fields, the conductivity components are 
\ba
\sigma_{yy} \!\!\!&=&\!\!\! -i\frac{r_H^2a^3_{y(2)}}{w_f\hspace{1mm}a^3_{y(0)}} \;, \;\;\; \;\;\; \;\;\;
\sigma_{xx} = -\frac{ir_H^2}{w_fa^3_{x(0)}}\left(a^3_{x(1)}+ {\Omega} \frac{i {w_f} a^2_{t(0)}+\mu a^1_{t(0)}}{\mu^2-w_f^2}\right).
\label{nn}
\ea
We show the numerical solution  for the real   and imaginary parts of $\sigma_{xx}$ and $\sigma_{yy}$ for the $z=2$ system in   figures 3.  As in the relativistic case  the conductivity components approach 1 at large $w_f$. We observe the formation of a gap in the real  part of $\sigma_{yy}$ as it happens in
the case of a Maxwell field coupled to a scalar \cite{horowitz} and in the purely Yang-Mills \cite{Gubser}  bulk Lagrangians cases. There is a pole in the imaginary parts of $\sigma_{xx}$ and $\sigma_{yy}$ at $w_f = 0$ characteristic of superconducting behavior. There is a second pole in the imaginary $\sigma_{xx}$ at $w_f = w^*_f =  0.199 \rho$ at $T/\rho = 0.022$ accompanied by the corresponding delta function in its  real part,  in agreement with Kramers-Kronig relations (this delta function is not represented in figure 3 left since the numerical procedure can only render continuous functions). The $w^*_f$ value obtained numerically satisfies $w^*_f = \mu $ as expected from eq.\eqref{nn}.
 This pole is absent in the analysis of \cite{Bu} for  a bulk Yang-Mills Lagrangian in the background of  a different $z=2$ Lifshitz black hole (the one presented in \cite{Taylor} with $g_2(u) = (1 - u^4)$  arising in the case in which the dilaton field is dynamical,  instead of the one we have used, eq.(\ref{greevy})). In \cite{Bu} such absence was attributed to the logarithmic behavior of $A_0$ resulting from  the $z=2$ scaling. Our result shows that  for the  $z=2$ black hole background that we used such logarithmic behavior does not prevent the existence
of this pole.

The analysis of the $z=4$ theory follows similarly and the behavior of conductivity components is qualitatively the same. We also find in this $z=4$ case, with $g_4(u) = 1 - u^4$, a second pole located at  $w_f^* = 20.5 \rho$ for
$T/\rho= 0.022$.

We shall end this work with a brief summary and a discussion of our results. We have studied  finite temperature effects in two models with different dynamical critical exponent using the gauge/gravity correspondence. Looking for a vector order parameter  and inspired by Gubser and Pufu's work
on $z=1$ $p$-wave holographic  superconductors   \cite{Gubser},  we have chosen as gravity dual a  Yang-Mills theory  in the gravitational background  of Lifshitz black-holes with $z=2$ and $z=4$. Apart from solving the equations of motion
in the bulk using a numerical approach, we have also extended the analytical
approximation developed in
\cite{Gregory}-\cite{Bellon} which allows to reproduce the numerical results  with remarkable
simplicity and precision.

~

\centerline{\includegraphics[width=1.1\textwidth]{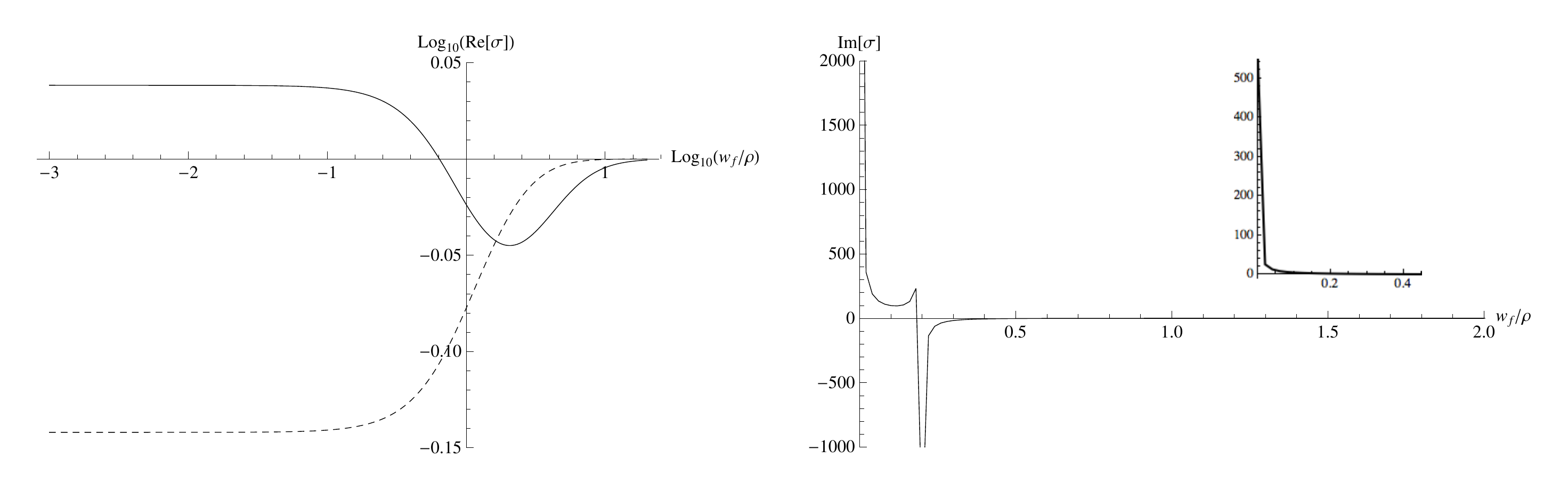}}

\noindent Figure 3:  Real an imaginary parts of conductivity as a function of the frequency
for $T/\rho= 0.022$ for the $z=2$ system. The solid line corresponds to $\sigma_{xx}$ and the dashed one to $\sigma_{yy}$.
The insert figure on the right displays a detail of the imaginary part of $\sigma_{yy}$  rendering visible the pole at $\omega_f = 0$.

~

 Although one could
 presume that the anisotropic scaling of the background metric would lead to a critical behavior    differing  from the one found in \cite{Gubser}  for $z=1$, our results show instead a remarkable resemblance with the relativistic case. In particular, the condensate has the typical $(T_z^c - T)^{1/2}$ mean field  behavior for $T$ close to the critical temperature $T^c_z$ both for $z=2$ and $z=4$.  The dependence on $z$ only affects the  coefficient in   the critical temperature  which grows with $z$, a behavior that could be argued to be valid
 for arbitrary $z$, as we illustrated applying our analytic approach to an heuristic black hole function $g(u;z)$ defined in eq.(\ref{estaforma}). Using the same approach we were able to extract  the
 condensate behavior  in the range of small temperatures where the probe approximation is valid, finding that   $\Omega \propto T^{-(z-2)/2z}$, in total agreement with the numerical calculations.
All these results confirm that the analytic approximation developed in
 \cite{Gregory}-\cite{Bellon} and refined here has proved to be sufficiently accurate as to avoid the necessity to resort to numerical methods.

 ~

\noindent\underline{Acknowledgments}:   This work was supported by  CONICET  , ANPCYT , CIC and UNLP, Argentina.

\end{document}